\documentclass{article}
\usepackage[utf8]{inputenc}
\usepackage{tabularx}
    \newcolumntype{L}{>{\centering\arraybackslash}X}
\usepackage{graphicx}
\usepackage{float}
\usepackage{setspace}
\usepackage{hyperref}
\usepackage[margin=1.5in]{geometry}
\usepackage[sorting=none,style=ieee]{biblatex}
\begin{document}
\title{Dataset of gold nanoparticle sizes and morphologies extracted from literature-mined microscopy images}

\author{Akshay Subramanian\textsuperscript{1,2}, Kevin Cruse\textsuperscript{2,3}, Amalie Trewartha\textsuperscript{2},\\ Xingzhi Wang\textsuperscript{2,4}, A. Paul Alivisatos\textsuperscript{2,3,4,5}, Gerbrand Ceder\textsuperscript{2,3,*}}   

\maketitle

\begin{flushleft}
\textbf{1} Indian Institute of Technology Roorkee, Roorkee, Uttarakhand 247667, India
\\
\textbf{2} Materials Sciences Division, Lawrence Berkeley National Laboratory, Berkeley, California 94720, United States
\\
\textbf{3} Department of Materials Science and Engineering, University of California, Berkeley, Berkeley, California 94720, United States
\\
\textbf{4} Department of Chemistry, University of California, Berkeley, Berkeley, California 94720, United States
\\
\textbf{5} Kavli Energy NanoScience Institute, Berkeley, California 94720, United States
\\
\textbf{*} Corresponding author: Gerbrand Ceder (gceder@berkeley.edu)
\end{flushleft}

\doublespacing

\begin{abstract}

    The factors controlling the size and morphology of nanoparticles have so far been poorly understood. Data-driven techniques are an exciting avenue to explore this field through the identification of trends and correlations in data. However, for these techniques to be utilized, large datasets annotated with the structural attributes of nanoparticles are required. While experimental SEM/TEM images collected from controlled experiments are reliable sources of this information, large-scale collection of these images across a variety of experimental conditions is expensive and infeasible. Published scientific literature, which provides a vast source of high-quality figures including SEM/TEM images, can provide a large amount of data at a lower cost if effectively mined. In this work, we develop an automated pipeline to retrieve and analyse microscopy images from gold nanoparticle literature and provide a dataset of 4361 SEM/TEM images of gold nanoparticles along with automatically extracted size and morphology information. The dataset can be queried to obtain information about the physical attributes of gold nanoparticles and their statistical distributions.

\end{abstract}

\section{Background and Summary}
\label{intro}
Scanning Electron Microscopy (SEM) and Transmission Electron Microscopy (TEM) are two of the most widely used imaging techniques for nanoparticles. A variety of qualitative and quantitative information about the structural characteristics of nanoparticles can be retrieved from the analysis of these images. In particular, information about the physical attributes of nanoparticles such as sizes and morphologies that have direct bearing on the properties of the material can be especially valuable. \cite{Mironava2010, Wani2013, Niikura2013, Chithrani2006} The factors controlling these attributes are poorly understood, and so aggregating this information on a large scale and identifying trends in this data is key to better understanding the synthesis and properties of nanoparticles. Data-driven and statistical analysis techniques can potentially play an important role in achieving this. \\ 

One possible avenue for acquisition of large numbers of images is through the published scientific literature, which provides a vast source of high-quality figures, including SEM/TEM images that can supply a large amount of data at a low cost if effectively mined. However, the complex structure and lack of uniformity in the way images are presented in literature makes the task of effectively retrieving and utilizing this information challenging. \cite{Kononova2021}\\

Over the last few years, there has been significant progress in the development of techniques to mine and analyse microscopy images present in scientific literature. Mukaddem et al. \cite{Mukaddem2020} developed ImageDataExtractor, a pipeline to automatically mine microscopy images from literature and perform statistical analysis of particle sizes. Kim et al. \cite{Kim2020} created a tool for the size analysis of nanoparticles in SEM images, distinguishing between two broad classes of nanoparticles: core-only and core-shell nanoparticles. Hiszpanski et al. \cite{Hiszpanski2020} utilized this tool to gather insights from nanoparticle literature. Schwenker et al. \cite{Schwenker} developed an automated pipeline to create a database of images extracted from literature that are 'self-labelled' with keywords retrieved from captions. \\

While a suite of techniques have been developed to extract microscopy images from literature and analyse particle sizes in them, there has been relatively less progress in development of techniques to identify nanoparticle morphologies (for example sphere, rod, cube etc.) at the particle-level in the extracted microscopy images. Morphology identification tools have been developed recently for experimental microscopy images using unsupervised clustering approaches \cite{Wang2021, Lee2020}, but they have not been designed to do so on more noisy images such as those from literature which contain artifacts and have high variability. Moreover, they have been developed as an additional step that is carried out after particle segmentation, which results in the accumulation of errors over the two steps. This lack in robust morphology identification tools is expected to have also contributed to the scarcity in publicly available morphology-annotated datasets of microscopy images. \\

In this work, we 
\begin{enumerate}
    \item develop a fully-automated, deep learning-based pipeline to retrieve and analyse microscopy images from scientific literature. The pipeline includes a unified Mask-RCNN \cite{He2017} model that can accurately segment nanoparticles (including overlapping particles), and simultaneously identify human-interpretable morphologies of the segmented particles at the particle-level. Through this we attempt to bridge the gap in morphology identification tools in past work. 
    \item provide a dataset of 4361 SEM/TEM images of gold nanoparticles along with automatically extracted size and morphology information. Through this, we attempt to address the need for publicly accessible morphology and size-annotated data of microscopy images. While the dataset described in this paper is a current snapshot, it will be updated dynamically as and when modifications are made.
\end{enumerate}

We believe that the importance of the techniques utilized in this work will be amplified as robotic synthesis of nanoparticles becomes used in the future, and can also be be extended/improved in several directions. One potential extension that may be possible is to analyse the correlations between synthesis recipes as described in the text \cite{textminingcruse}, and the synthesis outcomes (size, shape etc.) as extracted from images of scientific publications.

\section{Methods}

\subsection{Image Extraction Pipeline}
Our image extraction pipeline consists of six stages: 1) paper parsing, 2) sub-figure separation, 3) isolation of particulate microscopy images, 4) detection and interpretation of labels, scales and bars, 5) nanoparticles segmentation and morphology classification, and 6) size measurement. The techniques used in each stage are detailed in the following sections.

\subsubsection{Paper Parsing}
\label{parsing}
Parsers written using the Beautiful Soup Python library are used to automatically retrieve image URLs, captions and figure titles (for example, Figure 1 in this paper) from the HTML sources of papers. The scientific publications used in this work are drawn from the following four publishers: Elsevier, The Royal Society of Chemistry, Nature Publishing Group and Springer, each of which use their own HTML syntax/style. Hence, separate parsers had to be written for each publisher. In total, 131,103 articles were parsed in this stage. \\


These articles were obtained through a two-step natural language processing pipeline. First, from a database of 4,973,165 full-text materials science articles \cite{Kononova2019}, 811,905 nanomaterial articles were identified using regular expression matching for any word in the full text starting with "nano". Next, we used term frequency-inverse document frequency (TF-IDF) vectorization to represent the uniqueness of each word in each article, where every element of the TF-IDF vector of a given article represents the frequency of a word token normalized by the number of documents in which that word token appears across the entire corpus. For this step, we identified articles whose TF-IDF values for the words "gold" or "Au" were larger than any of "silver", "Ag", "copper", "Cu", "palladium", "Pd", "platinum", or "Pt", similar to the nanomaterial article processing pipeline developed in Hiszpanski et. al \cite{Hiszpanski2020}. This final step yielded the 130,692 gold nonmaterial articles used for TEM/SEM image extraction. \\


Once figures from all papers are parsed, those that are likely to contain SEM/TEM images of gold nanoparticles are selected with regular expression (regex) filters and then downloaded. We apply three categories of regex filters, based on: 1) type of microscopy image; 2) composition; and 3) morphology of particles. The keywords belonging to each category are shown in Table ~\ref{table:regex}. For a figure to be downloaded, we require its captions to match at least one keyword from each of the three categories. In addition to the keywords mentioned in the table, we also include grammatical variants of each by only matching for the stems of each keyword (stemming).
We also note that while we include "aunr" as a keyword, we do not include "aunp", "auns", "aunc" and "aunt" even though papers use them to refer to nanoparticles, nanospheres, nanocubes and nanotriangles respectively. This is because we observed that the same abbreviations are sometimes also used to refer to other morphologies. For example, aunc is used for gold nanocrystals, auns is also used for gold nanostars and aunp, which refers to gold nanoparticles, is used as a unified term that can refer to any morphology. So, including these terms results in the collection of microscopy images with undesired morphologies, and hence a loss in precision.
The downloaded images are carried forward to the next stage.

\subsubsection{Sub-figures Separation}
\begin{figure}[H]
  \centering
  \includegraphics[width=1\textwidth]{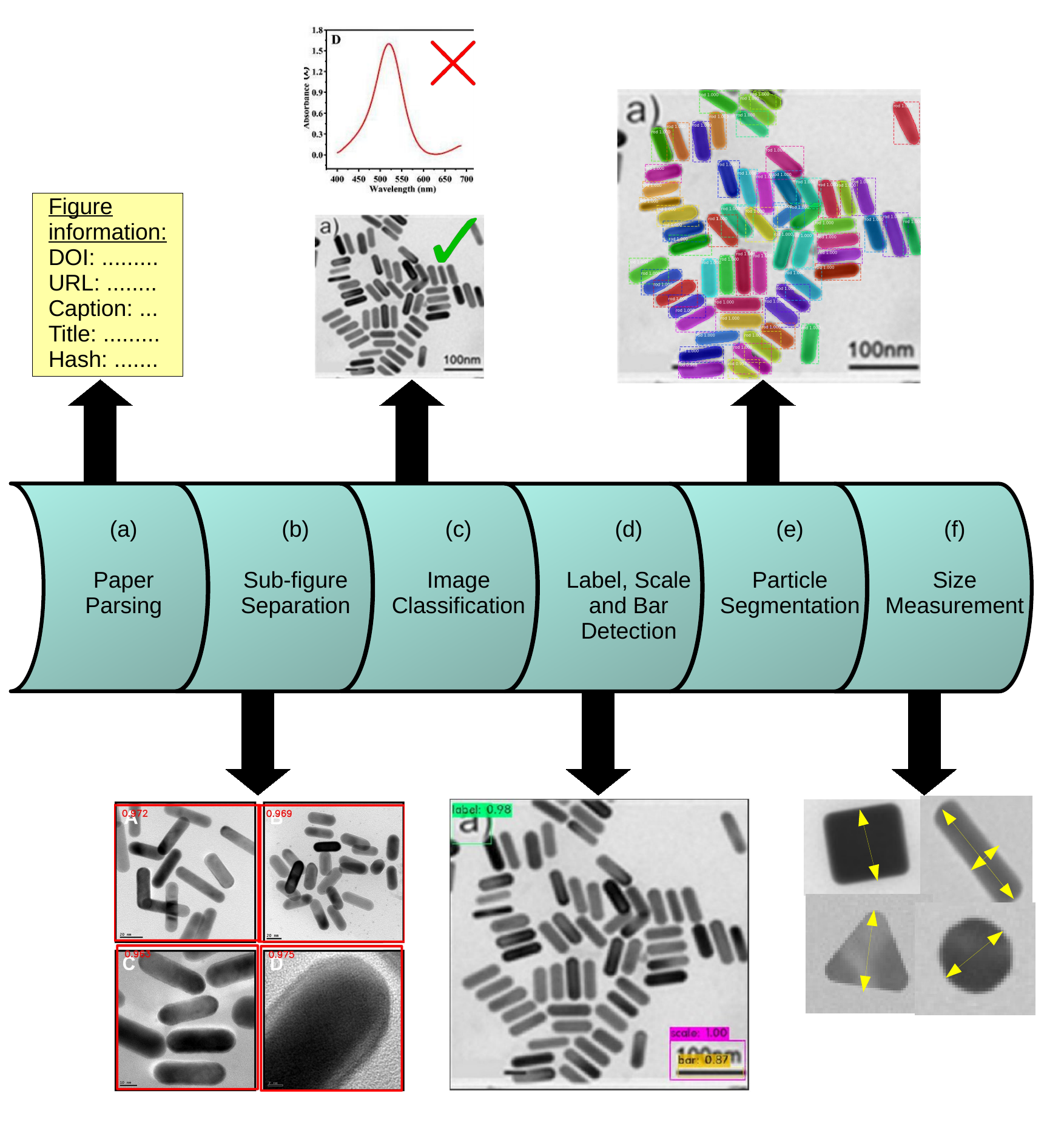}
  \caption{\textbf{Schematic representation of all major steps in the extraction pipeline.} (a) Information such as caption, DOI and Hash are parsed during the Paper Parsing stage. (b) Sub-figures are identified and separated from their parent composite figures. (c) Two binary classifiers are applied sequentially on sub-figures to isolate microscopy (SEM \& TEM) images that contain nanoparticles in them. (d) Labels, Scales and Bars are located and read with an OCR reader. (e) Nanoparticle segmentation and identification of particle morphologies is performed using a Mask-RCNN model. (f) Sizes of segmented particles are measured. The dimensions measured are chosen depending on the morphology of the particle. 
  Figures shown in sub-figures (b), (d) and (e) show actual predictions made by models from the pipeline. Numbers within/beside predictions in sub-figures (b) and (d) are the prediction confidences. The sample figure shown in (b) has been adapted with permission from ref. \cite{Jayabal2014}. Copyright 2014 Elsevier. The UV-vis spectrum shown in (c) has been adapted with permission from ref. \cite{Crulhas2016}. Copyright 2016 Springer Nature. The TEM image shown in (c) and (d) has been adapted with permission from ref. \cite{Si2018}. Copyright 2018 Elsevier. }
  \label{fig:pipeline}
\end{figure}

A majority (nearly 80\%) of the filtered images are composite figures. This makes the localization and separation of sub-figures a necessary step in the pipeline. For localization, we use the sub-figure localization algorithm that was developed by Tsutsui et al. \cite{Tsutsui2017}, which is based on the YOLOv2 object detection network \cite{Redmon2017}. Running this algorithm on the figures extracted in the previous stage gives us predicted sub-figure locations. These are then cropped out, separated and passed on to the Image Classification stage. Figure ~\ref{fig:pipeline}(b) shows a sample prediction of the sub-figure localization algorithm on an extracted image. \\

Two broad classes of approaches have been utilized in past literature for this task of sub-figure separation. The first class of approaches defines a sub-figure as an entity in a composite figure that can be uniquely identified by a label (i.e. a, b, c etc.), for example, work by Schwenker et al. \cite{Schwenker}. In their approach, labels in the composite figure are first localized and these locations are used to guide the detection of sub-figures. The second class of approaches simply treats any image present within the composite figure as a sub-figure, for example, work by Mukaddem et al. \cite{Mukaddem2020}. In their approach, the position of labels are not utilized in the process of sub-figure detection. \\

In cases where clusters of images fall under the same sub-figure label, the former approach is expected to identify the entire cluster as a single sub-figure, while the latter approach is expected to identify each image in the cluster as a separate sub-figure. In our work, we use the latter class of approaches, since our aim is to extract individual TEM/SEM images and clusters of them often fall under the same subfigure label. \\

An advantage of the first class of approaches however, is their inherent ability to match the separated subfigures with relevant portions of the caption. This task is significantly more challenging in the latter approach, specifically in cases where more than one image fall under the same label.

\subsubsection{Isolation of Particulate Microscopy Images}
This stage consists of two binary classifier models: Classifier-1 and Classifier-2, which are sequentially applied to the extracted sub-figures. Classifier-1 first identifies whether a given image is a microscopy (SEM/TEM) image or not. In the case it is identified as one, Classifier-2 then determines whether it contains nanoparticles or not. Through the sequential application of these classifiers on the entire database, we are able to isolate SEM/TEM images that contain nanoparticles in them. \\

Datasets to train and evaluate these classifiers were prepared by manually annotating a chosen set of sub-figures with relevant class labels. More details about the datasets used for each task are shown in Table ~\ref{table:classifier}. \\

We used the same model architecture and training procedure for both  classification tasks. A ResNet-50 \cite{He2016} deep learning model, which was pre-trained on the ImageNet dataset \cite{JiaDeng2009}, was fine-tuned on the training set for 500 epochs, and the best performing model on the validation set was chosen. The Stochastic Gradient Descent (SGD) optimizer was employed with learning rate of 6.5e-3 during training. To reduce the effects of imbalance in class frequencies on model training, we utilized weighted sampling of examples to ensure that the model ``saw" an equal number of examples from each class during every batch of training. Model preparation and training were done using the PyTorch deep learning framework \cite{Paszke2019} and tuning of hyperparameters was done using the RayTune framework \cite{Liaw2018}.

\subsubsection{Detection and Interpretation of Labels, Scales and Bars}
\label{label_scale_bar}
This stage consists of four major steps:
\begin{enumerate}
    \item A YOLOv4 \cite{Bochkovskiy2020} object detection network is used to locate labels, scales and bars in the SEM/TEM images shortlisted during the classification stage. The detected objects are then cropped and separated. A sample prediction is shown in Figure~\ref{fig:pipeline}(d).
    \item The cropped scales and labels are magnified and passed through an SRCNN super-resolution model \cite{Dong2016}.
    \item Labels and scales are read using the Tesseract \cite{Smith2007} OCR reader.
    \item The horizontal lengths (in pixels) of the cropped bars are then measured and mapped to the text (digit and unit) read from the corresponding scales.
\end{enumerate}

Since the SRCNN inference step (step 2) requires the cropped label/scale to be passed through a deep learning model, it is a relatively time-consuming step, and becomes a bottleneck if the label/scale is already clearly legible. Hence, we perform this step only if directly performing the OCR reading step (step 3) is unsuccessful. Success of the OCR reading step is determined by verifying if the read text satisfies certain criteria. For scales, this includes conditions such as (i) both a unit (nm or $\mu$m) as well as a number must be present in the read text and (ii) the read number should be divisible by 5 if it is greater than 10. We impose condition (ii) to improve the precision of read values since a large majority of scales tend to either be single digit numbers or double/triple digit numbers that are divisible by 5. For labels, the read text must consist of either a single letter or a letter followed by a digit in order to be considered successful. The Tesseract OCR reader tends to perform best when dark text is located on a light background.
Hence, for every failed read attempt of the OCR reader, we repeat the step after replacing the scale/label image with its color-inverted version. We observed that this improves the extraction recall by improving performance on light-colored scales and labels. \\

The YOLOv4 object detection network \cite{Bochkovskiy2020} mentioned in step 1 requires a labelled dataset to be prepared for model training. Details on the dataset can be found in Table \ref{table:classifier}. Annotation of ground truth bounding boxes and classes were performed using the LabelImg \cite{labelimg} software. Model training was carried out for 6000 iterations on the training set and the best performing model on the validation set was chosen. Training and evaluation were performed using the Darknet deep learning framework \cite{darknet13} and the model achieved a mean Average Precision (mAP@50) of 0.880 (0.963 on scales, 0.962 on labels, and 0.716 on bars) on the test set. The lower score on bars results from the high level of variation in the way that bars are presented in literature. For example, some papers use horizontal I-shaped lines, while others use lines calibrated with precision markers instead of bars. In order to reduce the effect of this score on performance, we use the horizontal length of the detected scale as an approximation of the bar length in cases where the model fails on bar detection but is successful in scale detection. Since bars are often present within scale boxes, we observed that lengths of scale boxes and bars are nearly equal in a majority of cases, and hence this can be considered a good approximation. \\

Past efforts on the task of label and scale detection have mostly used classical threshold-based segmentation techniques which make use of common attributes such as the rectangular shape and white color of label and scale boxes to identify them \cite{Mukaddem2020, Kim2020}. By instead using the YOLOv4 model for this task, we preclude any such prior assumptions on the size, shape, and color of scales and labels. Hence, unlike classical approaches that require separate rules to be written for labels, scales, and bars, the deep learning approach is able to detect all three objects in a single forward pass without the need for specification of any object-specific rules. Our approach is similar to the one used by Schwenker et al. \cite{Schwenker}.

\subsubsection{Nanoparticles Segmentation and Morphology Classification}
\label{segmentation}
For this task, we use the Mask-RCNN \cite{He2017}, an instance segmentation model that is capable of performing both the tasks of segmentation and object-level classification. We use its segmentation capability to perform nanoparticle segmentation in microscopy images and its classification capability to perform particle-level morphology identification. For the task of morphology identification, we consider only four morphologies in this work - sphere, rod, cube, and triangular prism since these are the shapes that occur most frequently in our database. Figure~\ref{fig:pipeline}(e) shows the prediction of the Mask-RCNN model on a sample image. \\

A dataset to train and evaluate the Mask-RCNN model was prepared by annotating microscopy images with segmentation masks and particle morphologies. More details on the dataset can be found in Table \ref{table:classifier}. Since each image contained nearly thirty particles on average, more than 5000 instances were annotated in total. All annotation was done using the LabelBox software \cite{labelbox}. We found it important to (i) include negative examples (images with no nanoparticles) and (ii) include both low and high resolution images in the training data, in order to improve the robustness of the trained model to the large variety of images observed in literature. We also included 131 experimental TEM images to increase the size of low frequency classes such as triangles. Of these 131 images, 114 are TEM images of gold nanoparticles that we synthesized \cite{Wang2021}, and 17 are microscopy images from the LLNL-MI-812379 release \cite{Kim2020}. We release this annotated dataset of 131 SEM/TEM images publicly on our GitHub repository. \\

For model training, ImageNet pre-trained weights were used as an initialization. Training was then carried out for 300 epochs, of which the first 50 involved training of only the network heads and the following 250 involved fine-tuning of the entire network. The model with the best performance on the validation set was chosen. The final model achieved a mean Average Precision (mAP@50) of 0.76 on the test set. While the segmentation and classification performances on all 4 morphologies were good, we did observe certain cases where cubes were mis-predicted as spheres, especially when the particles are very small in proportion to the size of the image. We also observed that the unique characteristics of nanoparticles as compared to objects present in standard computer vision tasks required certain hyperparameters to be tuned carefully. In particular, the sizes and aspect ratios of anchor boxes that are used in the Region Proposal Network were found to be the most important and needed to be tuned in accordance with the sizes and aspect ratios seen in TEM images of nanoparticles. In addition, we found data augmentation to significantly improve the performance of our model. The augmentation strategies utilized include vertical and horizontal flipping, rotation and Gaussian blurring of images. By augmenting the training data with these transformed versions of the images, we improved the robustness of the model to various orientations and degrees of noise in SEM/TEM images. We believe that this is especially helpful in our case since the size of the training dataset is relatively small. \\

Past work has primarily used classical image processing techniques for the task of nanoparticle segmentation, such as Thresholding, Hough transforms and the Watershed algorithm. \cite{Groom2018, Meng2018, Mirzaei2017} These techniques typically require a significant amount of user-tuned parameters. A drawback with these classical techniques is their inability to separate overlapping particles when the degree of overlap is high, which is a very common characteristic of TEM/SEM images containing nanoparticles. By instead using the Mask-RCNN, we significantly improve performance on such cases. Figure ~\ref{fig:violin}(a) shows a comparison of performances shown by a "classical" algorithm (watershed algorithm) and the Mask-RCNN on a TEM image from our database. While the classical approach fails to distinguish between particles that overlap, the Mask-RCNN model is clearly able to classify them as separate instances. The implementation of the Watershed algorithm used is available on our GitHub repository. \\

While there have been recent approaches \cite{Zhang2019, Frei2020, Wu2020, Yildirim2021} that have also utilized deep learning approaches to the task of nanoparticle segmentation and thereby perform well on overlapping particles, we believe a key feature of our approach using the Mask-RCNN model is its added ability to identify particle morphologies. In addition, by incorporating both segmentation and morphology identification into a single forward pass, we prevent errors from being accumulated between the two tasks.

\subsection{Size Measurement}
\label{size_measurement}
We measure sizes differently for particles belonging to different morphologies. A common initial step for all morphologies is to locate the centroid of the particle. We then simulate lines originating from the centroid and intersecting the boundaries of the particle. Constraints are then imposed on these lines in order to shortlist those that span the dimensions we are interested in measuring. For example, in order to measure the length of a rod, we would simply choose the longest line (which would connect the centroid of the rod to the rod tip) and double its length. Similarly, the width can be obtained by doubling the length of the shortest line.
The dimensions measured for each morphology are:
1) rod: length and width,
2) cube: side length,
3) triangular prism: height, and 
4) sphere: diameter (averaged over all diameters).
These measured dimensions are then converted into the unit that has been read from the scale to arrive at the final magnitude of the dimension.

\subsection{Web Application}
\begin{figure}[H]
  \centering
  \includegraphics[width=\textwidth]{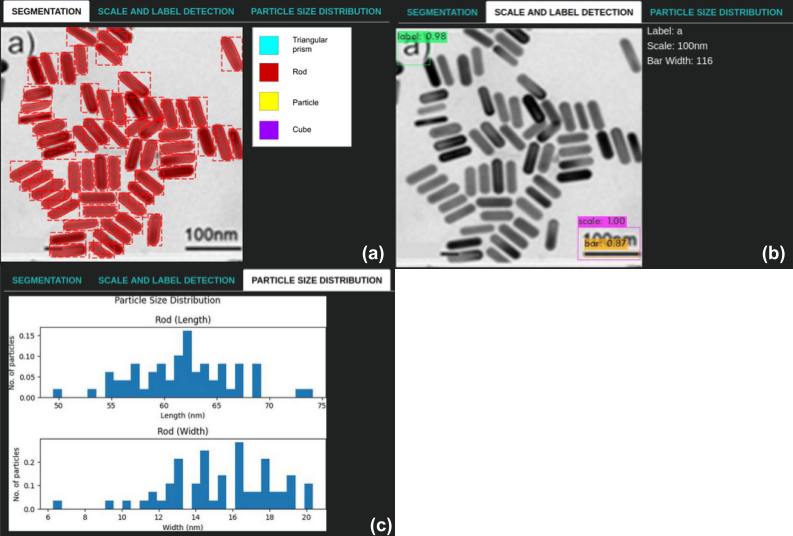}
  \caption{ \textbf{Illustration of the results table displayed on the web application when a sample TEM image is uploaded.} (a) Segmentation tab (b) Scale and Label Detection tab (c) Particle Size Distribution tab. Separate Size Distribution histograms are plotted for each measured dimension for all morphologies identified in the image. The sample image shown in (a) and (b) has been adapted with permission from ref. \cite{Si2018}. Copyright 2018 Elsevier.
}
  \label{fig:web_app}
\end{figure}

\begin{figure}[H]
  \centering
  \includegraphics[width=1.0\textwidth]{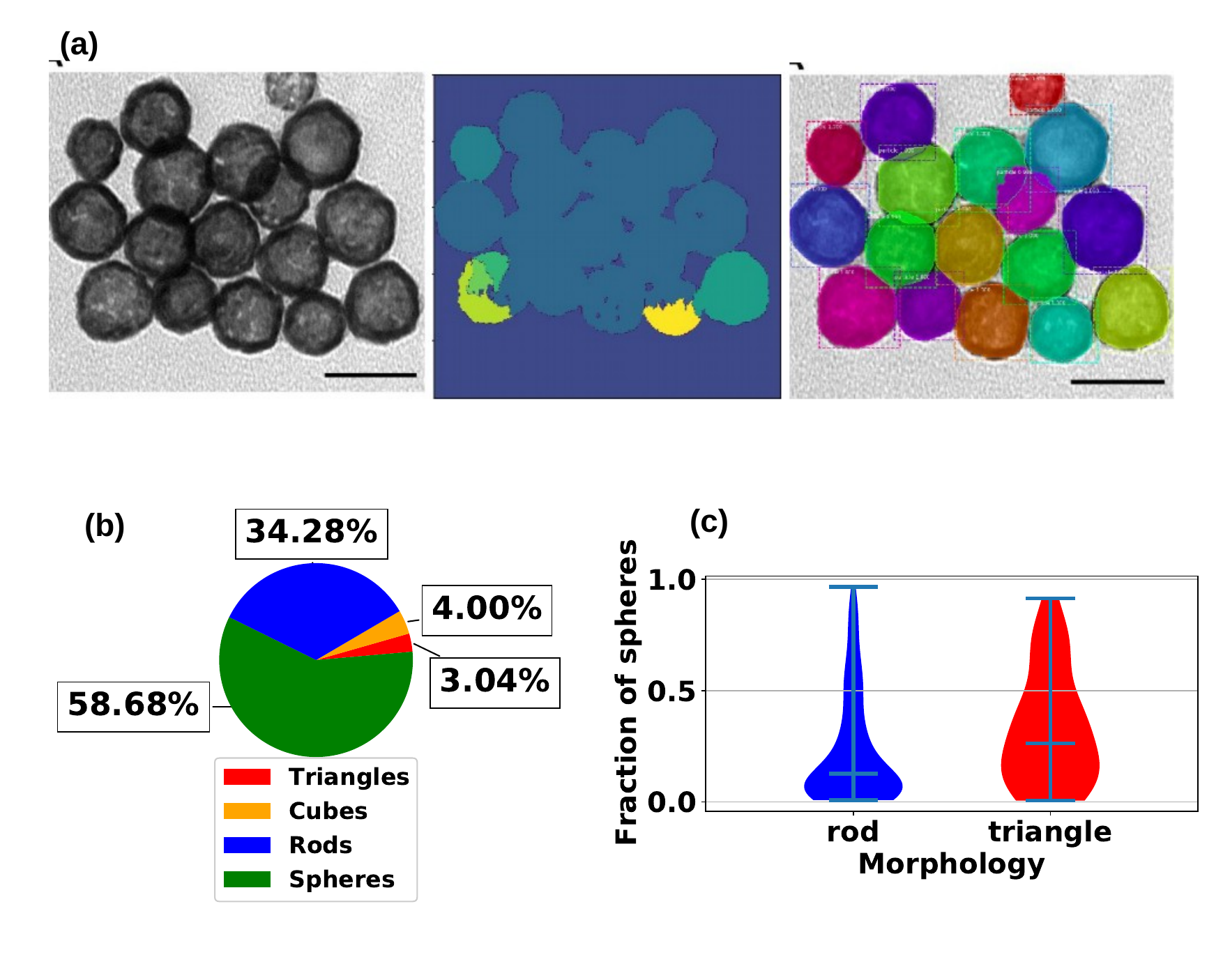}
  \caption{(a) Figure illustrating a comparison between the Watershed segmentation and the Mask-RCNN algorithms on segmentation of overlapping nanoparticles. The original TEM image is shown on the left, the results of the Watershed algorithm are shown in the middle, and the results of the Mask-RCNN are shown on the right. The image shown has been adapted with permission from ref. \cite{Lu2010}. Copyright 2010 Elsevier (b) Pie-chart showing fractional presence of various morphologies in the extracted dataset. The percentage shown beside each section can be interpreted as the percent of microscopy images in the dataset that contain a particular morphology in majority. Sections corresponding to spheres, triangles, rods and cubes represent 2491, 129, 1455 and 170 data points respectively. (c) Violin plots showing the fraction of spheres that co-occur with rods and triangular prisms. Horizontal lines intersecting the plots mark the positions of the medians and extrema. The plot corresponding to rod (blue) represents 325 data points while the plot corresponding to triangle (red) represents 36 data points. }
  \label{fig:violin}
\end{figure}

To illustrate the capabilities of the described image extraction and analysis techniques, we have developed a web application for automated analysis of TEM and SEM images. This allows users to easily experiment with our algorithms on custom images using a Graphical User Interface (GUI). The web application supports all image analysis techniques mentioned so far and additionally allows users to view statistics such as particle size distributions. It is accessible at \href{https://github.com/CederGroupHub/TEMExtraction-webapp}{https://github.com/CederGroupHub/TEMExtraction-webapp}. Figure ~\ref{fig:web_app} illustrates the analysis table displayed on the web application when a custom microscopy image is uploaded.\\

\section{Data Records}

The complete dataset consists of 4361 literature-mined microscopy (SEM and TEM) images of gold nanoparticles along with extracted particle sizes, morphologies, and metadata corresponding to each image. Metadata includes fields such as the DOI of the parent paper, URL of the parent composite figure, and a unique hash for each image. \\

The dataset is publicly available as a JSON file (\href{https://doi.org/10.6084/m9.figshare.17019836.v2}{https://doi.org/10.6084/\\m9.figshare.17019836.v2}) and an associated Python script (\href{https://github.com/CederGroupHub/TEMExtraction}{https://github.com/\\CederGroupHub/TEMExtraction}). The JSON file contains a mapping between image names and their corresponding particle sizes, morphologies, and image metadata (complete schema shown in Table ~\ref{table:full_schema}). The associated Python script can be used to download all the microscopy images onto a local system. It achieves this by 1) Retrieving and downloading all parent composite figures from the URLs present in the JSON file 2) Separating sub-figures from composite figures, and 3) Isolating the subset of images that belong to the dataset. \\

The set of papers that were used in this work were shortlisted from a much larger set of Materials Science papers. More details on how this was achieved are given in Section \ref{parsing}.


\section{Technical Validation}

\subsection{Extraction Accuracy}
All major steps in the pipeline have metric scores greater than 80\%, as can be seen in Table \ref{table:classifier}. \\ 

We noted that the recall of the OCR reading step (step 2 in Section \ref{label_scale_bar}) was particularly low (30-40\%), i.e., 
only 30-40\% of the scales and labels in microscopy images are successfully read by the OCR reader. We observed that this low recall arises from the inability of the OCR reader to perform "successfully" (see Section \ref{label_scale_bar} for definition of "success") on cropped labels/scales having low resolution or low contrast between text and background (black text on a dark background for example). A high percentage of cropped labels/scales have low resolutions since they often occupy a very small portion of the parent microscopy image.

\subsection{Morphology Distribution}
Figure ~\ref{fig:violin}(b) illustrates the distribution of images among the four morphologies. The percentages shown beside each section refer to the fractional percentage of images that have a majority of particles belonging to a given morphology. For example, 58.68\% of the images have spheres as their majority morphology. The pie chart indicates that there is a large imbalance in the morphologies present in literature images, with spheres being most common, followed by rods, cubes and finally triangles.

\subsection{Impurities}
Each SEM/TEM image can contain nanoparticles of one or more morphology classes. When we explore the dataset to try and identify trends in the co-occurrence of various morphologies, we observe that spheres co-occur with other morphologies most often. This observation matches with intuition since spheres are often formed as a byproduct during the shape controlled synthesis of more unique nanoparticle morphologies such as gold nanorods. \cite{Ha2007, Vigderman2012, Ye2013, Wang2021}\\

We then explored the fractional distribution of co-occurring morphologies in each image. The fraction of spheres in an image can be treated as an inverse measure of purity of synthesis since spheres are often the undesired byproducts that are formed in the synthesis of a target morphology. Figure ~\ref{fig:violin}(c) illustrates the distributions of sphere fractions that rods and triangles co-occur with. We have not included a violin plot for cubes, as the performance of the morphology classification step was relatively low for cubes, in particular in distinguishing cubes and spheres. It can be seen that the majority of images that have triangles or rods in them have less than 40\% of spherical impurities, and the median fractions of spherical impurities fall below 0.3 in both cases. Thus, our method can potentially serve as an objective measure of the purity of the product of a synthesis. The development of such a measure would allow future researchers to compare the shape purity of a synthesis to those found in existing literature, enabling them to evaluate the quality of a newly developed synthetic method.

\subsection{Size/Shape Distribution}
\label{size_shape_dist}
\begin{figure}[H]
  \centering
  \includegraphics[width=1.0\textwidth]{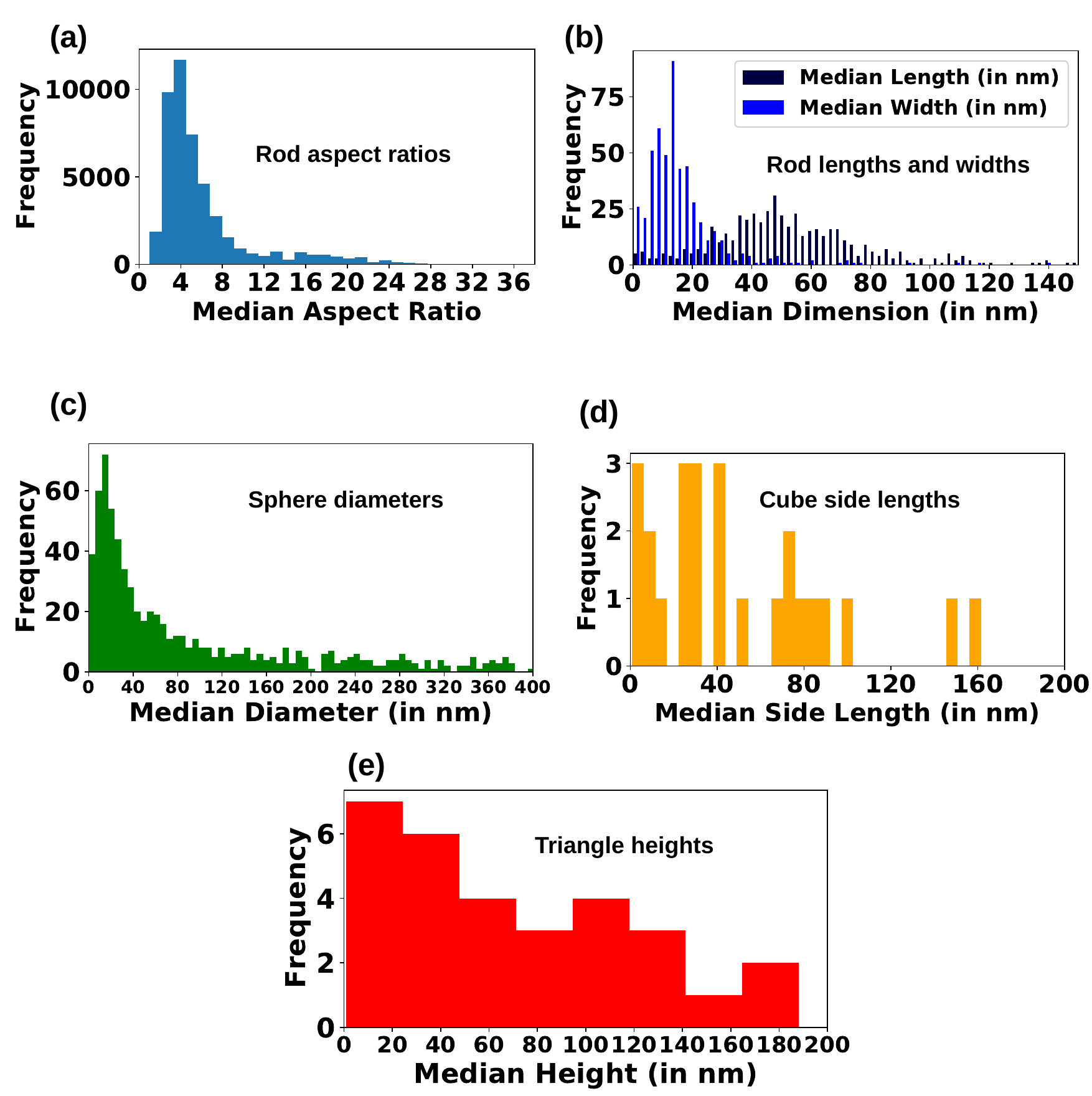}
  \caption{\textbf{Size distribution histograms for various dimensions of rods, cubes, triangles and spheres.} Distributions of (a) Aspect Ratios of rods (b) Lengths and Widths of rods (c) Diameters of spheres (d) Side Lengths of cubes (e) Heights of triangular prisms in the dataset. Median sizes from each microscopy image have been used in order to minimize the effects of outliers/mis-predictions which may have very large/small sizes in comparison to other particles in the image. x-axes of histograms (b), (c) and (d) have been truncated to make the peak positions more clearly visible, since peak positions are more reliable than the range/spread (More details in Section \ref{size_shape_dist}). }
  \label{fig:size_dist}
\end{figure}

We next looked at the distribution of particle sizes in our dataset. Figure \ref{fig:size_dist} shows size distribution histograms for each of the four morphologies. The measured dimensions are chosen depending on the morphology (as detailed in Section  \ref{size_measurement}). \\

We note that the peak position in each histogram is in general more reliable than the range/spread. This is because of errors in the pipeline that are introduced because of the presence of certain special classes of microscopy images in the dataset. These errors lead to noise in values in the histograms corresponding to very high or very low sizes. In particular, TEM/SEM figures sometimes have zoomed-in/magnified images embedded within them, which also tend to have their own scales and bars. This class of figures are often used when authors need to describe characteristics of nanoparticles at both coarse and fine levels. A special characteristic of these figures is that multiple magnifications of nanoparticles and multiple scales/bars exist in a single image. The current version of our pipeline is unable to differentiate between particles of embedded and parent figures and also between scales of embedded and parent figures. We simply measure the sizes of all particles in an image and perform unit conversion based on the value read from the single scale that has been identified with the highest confidence by the Label, Scale and Bar Detector. Hence, depending on which scale is chosen by the detector, sizes of some particles in this class of images will be overestimated or underestimated. \\

It is also interesting to note that dimensionless quantities are unaffected by this issue of over/underestimation, since they remain unaffected by changes in magnification and scale readings. Aspect ratio is one such example. Figure \ref{fig:size_dist} (a) shows the distribution of aspect ratios of rods in the dataset. 
For such cases, both the peak position as well as the range/spread of the histogram can be treated as reliable measures of what is seen in literature. Another property of dimensionless quantities is that they are unaffected by the errors introduced in the scale detection and OCR reading steps. Hence, aspect ratio measurements, for instance, can be carried out for all images in the dataset, even in cases where scales are not identified in them. This is why Figure ~\ref{fig:size_dist}(a) has higher frequencies than the other three histograms.


\section{Usage Notes}
The full dataset is available publicly at as a JSON file and an associated Python script. The latter is required to download the microscopy images on a local system. The annotated training dataset used for training the Mask-RCNN model (see Section \ref{segmentation}) is also available publicly in the same GitHub repository in a similar format (JSON and accompanied Python script). Detailed instructions to install all required libraries are provided in the README. The JSON file can be queried using a language of choice (for example Python, Matlab, R and Wolfram Mathematica).

\section{Code Availability}
\label{code}
All code used to retrieve and analyse microscopy images from literature is publicly available at \href{https://github.com/CederGroupHub/TEMExtraction}{https://github.com/CederGroupHub/TEMExtraction}, and instructions to install all required dependencies and run the pipeline are detailed in the README. All machine learning frameworks used are open source: Tensorflow \cite{tensorflow2015-whitepaper}, Keras \cite{chollet2015keras}, PyTorch \cite{Paszke2019} and Darknet \cite{darknet13}.

\section{Acknowledgements}
This work was funded by the U.S. Department of Energy, Office of Science, Office of Basic Energy Sciences, Materials Sciences and Engineering Division under Contract No. DE-AC02-05-CH11231 (D2S2 program KCD2S2). We thank Timothy Vollmer, Anna Sackmann and Rachael Samberg (Science Data and Engineering Librarians at UC Berkeley) for helping us obtain Text and Data Mining agreements with the specified publishers. We also thank Maria Chan, Weixin Jiang and other group members for the valuable discussions on compound figure separation techniques, as well as Sam Gleason, Jakob Dahl, Caitlin McCandler and other members of the Alivisatos, Persson, Jain and Sutter-Fella groups for insightful discussions about gold nanoparticle synthesis and advise on analysis of the data.

\section{Author Contributions}
A.S. developed the image extraction pipeline, nanoparticle size and shape identification algorithms, and analyzed the data. K.C. developed the text mining pipeline to isolate gold nanoparticles papers from the larger set of materials science papers. A.T. provided guidance on the development of the pipeline, size/shape identification algorithms and data analysis. X.W. generated experimental TEM images as training data for the segmentation model. P.A. supervised the project. G.C. developed the approach, and supervised the project. All authors discussed the results and wrote the final manuscript.

\section{Competing Interests}
The authors declare no competing interests.

\printbibliography[heading=bibnumbered]

@article{Lu2010,
author = {Lu, Wei and Huang, Qian and Ku, Geng and Wen, Xiaoxia and Zhou, Min and Guzatov, Dmitry and Brecht, Peter and Su, Richard and Oraevsky, Alexander and Wang, Lihong V. and Li, Chun},
doi = {10.1016/j.biomaterials.2009.12.007},
issn = {01429612},
journal = {Biomaterials},
number = {9},
pages = {2617--2626},
pmid = {20036000},
publisher = {Elsevier Ltd},
title = {{Photoacoustic imaging of living mouse brain vasculature using hollow gold nanospheres}},
url = {http://dx.doi.org/10.1016/j.biomaterials.2009.12.007},
volume = {31},
year = {2010}
}

@article{Si2018,
author = {Si, Yuelei and Cao, Shuang and Wu, Zhijiao and Ji, Yinglu and Mi, Yang and Wu, Xiaochun and Liu, Xinfeng and Piao, Lingyu},
doi = {10.1016/j.apcatb.2017.08.024},
issn = {09263373},
journal = {Applied Catalysis B: Environmental},
pages = {471--476},
publisher = {Elsevier B.V.},
title = {{What is the predominant electron transfer process for Au NRs/TiO2 nanodumbbell heterostructure under sunlight irradiation?}},
url = {http://dx.doi.org/10.1016/j.apcatb.2017.08.024},
volume = {220},
year = {2018}
}

@article{Jayabal2014,
author = {Jayabal, Subramaniam and Ramaraj, Ramasamy},
doi = {10.1016/j.apcata.2013.10.056},
issn = {0926860X},
journal = {Applied Catalysis A: General},
pages = {369--375},
publisher = {Elsevier B.V.},
title = {{Bimetallic Au/Ag nanorods embedded in functionalized silicate sol-gel matrix as an efficient catalyst for nitrobenzene reduction}},
url = {http://dx.doi.org/10.1016/j.apcata.2013.10.056},
volume = {470},
year = {2014}
}

@article{Crulhas2016,
author = {Crulhas, Bruno P. and Ramos, Naira P. and Castro, Gustavo R. and Pedrosa, Valber A.},
doi = {10.1007/s10008-016-3182-y},
issn = {14328488},
journal = {Journal of Solid State Electrochemistry},
number = {9},
pages = {2427--2433},
title = {{Detection of hydrogen peroxide releasing from prostate cancer cell using a biosensor}},
volume = {20},
year = {2016}
}

@misc{textminingcruse,
  title={{Text-mined AuNP Synthesis Recipes Dataset}},
  author={Cruse, Kevin and Trewartha, Amalie and Lee, Sanghoon and Wang, Zheren and Huo, Haoyan and He, Tanjin and Kononova, Olga and Jain, Anubhav and Ceder, Gerbrand},
  year={2021},
  publisher={figshare},
  note={Dataset},
  howpublished={\url{https://doi.org/10.6084/m9.figshare.16614262.v3}},
}

@misc{chollet2015keras,
  title={Keras},
  author={Chollet, Fran\c{c}ois and others},
  year={2015},
  publisher={GitHub},
  howpublished={\url{https://github.com/keras-team/keras}},
}

@misc{tensorflow2015-whitepaper,
title={ {TensorFlow}: Large-Scale Machine Learning on Heterogeneous Systems},
url={https://www.tensorflow.org/},
note={Software available from tensorflow.org},
author={
    Mart{\i}n~Abadi and
    Ashish~Agarwal and
    Paul~Barham and
    Eugene~Brevdo and
    Zhifeng~Chen and
    Craig~Citro and
    Greg~S.~Corrado and
    Andy~Davis and
    Jeffrey~Dean and
    Matthieu~Devin and
    Sanjay~Ghemawat and
    Ian~Goodfellow and
    Andrew~Harp and
    Geoffrey~Irving and
    Michael~Isard and
    Yangqing Jia and
    Rafal~Jozefowicz and
    Lukasz~Kaiser and
    Manjunath~Kudlur and
    Josh~Levenberg and
    Dandelion~Man\'{e} and
    Rajat~Monga and
    Sherry~Moore and
    Derek~Murray and
    Chris~Olah and
    Mike~Schuster and
    Jonathon~Shlens and
    Benoit~Steiner and
    Ilya~Sutskever and
    Kunal~Talwar and
    Paul~Tucker and
    Vincent~Vanhoucke and
    Vijay~Vasudevan and
    Fernanda~Vi\'{e}gas and
    Oriol~Vinyals and
    Pete~Warden and
    Martin~Wattenberg and
    Martin~Wicke and
    Yuan~Yu and
    Xiaoqiang~Zheng},
  year={2015},
}

@misc{labelimg,
    title={LabelImg},
    year={2015},
    url={https://github.com/tzutalin/labelImg}
}

@misc{darknet13,
  author =   {Joseph Redmon},
  title =    {Darknet: Open Source Neural Networks in C},
  howpublished = {\url{http://pjreddie.com/darknet/}},
  year = {2013}
}

@misc{labelbox,
    title={LabelBox},
    year={2021},
    url={https://labelbox.com}
}

@article{Ha2007,
author = {Ha, Tai Hwan and Koo, Hee Joon and Chung, Bong Hyun},
doi = {10.1021/jp066454l},
issn = {19327447},
journal = {Journal of Physical Chemistry C},
number = {3},
pages = {1123--1130},
title = {{Shape-controlled syntheses of gold nanoprisms and nanorods influenced by specific adsorption of halide ions}},
volume = {111},
year = {2007}
}

@article{Vigderman2012,
author = {Vigderman, Leonid and Khanal, Bishnu P. and Zubarev, Eugene R.},
doi = {10.1002/adma.201201690},
issn = {09359648},
journal = {Advanced Materials},
number = {36},
pages = {4811--4841},
pmid = {22740090},
title = {{Functional gold nanorods: Synthesis, self-assembly, and sensing applications}},
volume = {24},
year = {2012}
}

@article{Ye2013,
author = {Ye, Xingchen and Zheng, Chen and Chen, Jun and Gao, Yuzhi and Murray, Christopher B.},
doi = {10.1021/nl304478h},
issn = {15306984},
journal = {Nano Letters},
number = {2},
pages = {765--771},
title = {{Using binary surfactant mixtures to simultaneously improve the dimensional tunability and monodispersity in the seeded growth of gold nanorods}},
volume = {13},
year = {2013}
}

@article{Chithrani2006,
author = {Chithrani, B. Devika and Ghazani, Arezou A. and Chan, Warren C.W.},
doi = {10.1021/nl052396o},
issn = {15306984},
journal = {Nano Letters},
number = {4},
pages = {662--668},
pmid = {16608261},
title = {{Determining the size and shape dependence of gold nanoparticle uptake into mammalian cells}},
volume = {6},
year = {2006}
}

@article{Kononova2021,
author = {Kononova, Olga and He, Tanjin and Huo, Haoyan and Trewartha, Amalie and Olivetti, Elsa A. and Ceder, Gerbrand},
doi = {10.1016/j.isci.2021.102155},
issn = {25890042},
journal = {iScience},
number = {3},
pages = {102155},
publisher = {Elsevier Inc.},
title = {{Opportunities and challenges of text mining in aterials research}},
url = {https://doi.org/10.1016/j.isci.2021.102155},
volume = {24},
year = {2021}
}

@article{Niikura2013,
author = {Niikura, Kenichi and Matsunaga, Tatsuya and Suzuki, Tadaki and Kobayashi, Shintaro and Yamaguchi, Hiroki and Orba, Yasuko and Kawaguchi, Akira and Hasegawa, Hideki and Kajino, Kiichi and Ninomiya, Takafumi and Ijiro, Kuniharu and Sawa, Hirofumi},
doi = {10.1021/nn3057005},
issn = {19360851},
journal = {ACS Nano},
number = {5},
pages = {3926--3938},
pmid = {23631767},
title = {{Gold nanoparticles as a vaccine platform: Influence of size and shape on immunological responses in vitro and in vivo}},
volume = {7},
year = {2013}
}

@article{Mironava2010,
author = {Mironava, Tatsiana and Hadjiargyrou, Michael and Simon, Marcia and Jurukovski, Vladimir and Rafailovich, Miriam H.},
doi = {10.3109/17435390903471463},
issn = {17435390},
journal = {Nanotoxicology},
number = {1},
pages = {120--137},
pmid = {20795906},
title = {{Gold nanoparticles cellular toxicity and recovery: Effect of size, concentration and exposure time}},
volume = {4},
year = {2010}
}

@article{Wani2013,
author = {Wani, Irshad A. and Ahmad, Tokeer},
doi = {10.1016/j.colsurfb.2012.06.005},
issn = {09277765},
journal = {Colloids and Surfaces B: Biointerfaces},
pages = {162--170},
pmid = {22796787},
publisher = {Elsevier B.V.},
title = {{Size and shape dependant antifungal activity of gold nanoparticles: A case study of Candida}},
url = {http://dx.doi.org/10.1016/j.colsurfb.2012.06.005},
volume = {101},
year = {2013}
}

@article{Zhang2019,
author = {Zhang, Fang and Zhang, Qian and Xiao, Zhitao and Wu, Jun and Liu, Yanbei},
doi = {10.1145/3373509.3373590},
isbn = {9781450376570},
journal = {ACM International Conference Proceeding Series},
pages = {205--212},
title = {{Spherical nanoparticle parameter measurement method based on mask r-cnn segmentation and edge fitting}},
year = {2019}
}

@article{Frei2020,
archivePrefix = {arXiv},
arxivId = {1907.05112},
author = {Frei, M. and Kruis, F. E.},
doi = {10.1016/j.powtec.2019.10.020},
eprint = {1907.05112},
issn = {1873328X},
journal = {Powder Technology},
pages = {324--336},
publisher = {Elsevier B.V.},
title = {{Image-based size analysis of agglomerated and partially sintered particles via convolutional neural networks}},
url = {https://doi.org/10.1016/j.powtec.2019.10.020},
volume = {360},
year = {2020}
}

@article{Wu2020,
author = {Wu, Yuanyi and Lin, Mengxing and Rohani, Sohrab},
doi = {10.1016/j.cherd.2020.03.004},
issn = {02638762},
journal = {Chemical Engineering Research and Design},
number = {Ld},
pages = {114--125},
publisher = {Institution of Chemical Engineers},
title = {{Particle characterization with on-line imaging and neural network image analysis}},
url = {https://doi.org/10.1016/j.cherd.2020.03.004},
volume = {157},
year = {2020}
}

@article{Groom2018,
author = {Groom, D. J. and Yu, K. and Rasouli, S. and Polarinakis, J. and Bovik, A. C. and Ferreira, P. J.},
doi = {10.1016/j.ultramic.2018.06.002},
issn = {18792723},
journal = {Ultramicroscopy},
number = {June},
pages = {25--34},
pmid = {30056278},
title = {{Automatic segmentation of inorganic nanoparticles in BF TEM micrographs}},
volume = {194},
year = {2018}
}

@article{Lee2020,
author = {Lee, Byoungsang and Yoon, Seokyoung and Lee, Jin Woong and Kim, Yunchul and Chang, Junhyuck and Yun, Jaesub and Ro, Jae Chul and Lee, Jong Seok and Lee, Jung Heon},
doi = {10.1021/acsnano.0c06809},
issn = {1936086X},
journal = {ACS Nano},
number = {12},
pages = {17125--17133},
title = {{Statistical Characterization of the Morphologies of Nanoparticles through Machine Learning Based Electron Microscopy Image Analysis}},
volume = {14},
year = {2020}
}

@article{Wang2021,
author = {Wang, Xingzhi and Li, Jie and Ha, Hyun Dong and Dahl, Jakob C. and Ondry, Justin C. and Moreno-Hernandez, Ivan and Head-Gordon, Teresa and Alivisatos, A. Paul},
doi = {10.1021/jacsau.0c00030},
issn = {2691-3704},
journal = {JACS Au},
number = {3},
pages = {316--327},
title = {{AutoDetect-mNP: An Unsupervised Machine Learning Algorithm for Automated Analysis of Transmission Electron Microscope Images of Metal Nanoparticles}},
volume = {1},
year = {2021}
}

@article{Meng2018,
author = {Meng, Yingchao and Zhang, Zhongping and Yin, Huaqiang and Ma, Tao},
doi = {10.1016/j.micron.2017.12.002},
issn = {09684328},
journal = {Micron},
number = {August 2017},
pages = {34--41},
pmid = {29304431},
publisher = {Elsevier},
title = {{Automatic detection of particle size distribution by image analysis based on local adaptive canny edge detection and modified circular Hough transform}},
url = {https://doi.org/10.1016/j.micron.2017.12.002},
volume = {106},
year = {2018}
}

@article{Mirzaei2017,
author = {Mirzaei, Mohsen and Rafsanjani, Hossein Khodabakhshi},
doi = {10.1016/j.micron.2017.02.008},
issn = {09684328},
journal = {Micron},
pages = {86--95},
pmid = {28282550},
publisher = {Elsevier Ltd},
title = {{An automatic algorithm for determination of the nanoparticles from TEM images using circular hough transform}},
url = {http://dx.doi.org/10.1016/j.micron.2017.02.008},
volume = {96},
year = {2017}
}

@article{Hiszpanski2020,
author = {Hiszpanski, Anna M. and Gallagher, Brian and Chellappan, Karthik and Li, Peggy and Liu, Shusen and Kim, Hyojin and Han, Jinkyu and Kailkhura, Bhavya and Buttler, David J. and Han, Thomas Yong Jin},
doi = {10.1021/acs.jcim.0c00199},
issn = {15205142},
journal = {Journal of Chemical Information and Modeling},
number = {6},
pages = {2876--2887},
pmid = {32286818},
title = {{Nanomaterial Synthesis Insights from Machine Learning of Scientific Articles by Extracting, Structuring, and Visualizing Knowledge}},
volume = {60},
year = {2020}
}

@article{JiaDeng2009,
author = {{Jia Deng} and {Wei Dong} and Socher, R. and {Li-Jia Li} and {Kai Li} and {Li Fei-Fei}},
doi = {10.1109/cvprw.2009.5206848},
isbn = {9781424439911},
pages = {248--255},
publisher = {IEEE},
title = {{ImageNet: A large-scale hierarchical image database}},
year = {2009}
}

@article{Tsutsui2017,
archivePrefix = {arXiv},
arxivId = {1703.05105},
author = {Tsutsui, Satoshi and Crandall, David J.},
doi = {10.1109/ICDAR.2017.93},
eprint = {1703.05105},
isbn = {9781538635865},
issn = {15205363},
journal = {Proceedings of the International Conference on Document Analysis and Recognition, ICDAR},
pages = {533--540},
title = {{A Data Driven Approach for Compound Figure Separation Using Convolutional Neural Networks}},
volume = {1},
year = {2017}
}

@article{He2017,
author = {He, Kaiming and Gkioxari, Georgia and Dollar, Piotr and Girshick, Ross},
doi = {10.1109/ICCV.2017.322},
isbn = {9781538610329},
issn = {15505499},
journal = {Proceedings of the IEEE International Conference on Computer Vision},
pages = {2980--2988},
title = {{Mask R-CNN}},
volume = {2017-Octob},
year = {2017}
}

@article{Yildirim2021,
author = {Yildirim, Batuhan and Cole, Jacqueline M.},
doi = {10.1021/acs.jcim.0c01455},
issn = {15205142},
journal = {Journal of Chemical Information and Modeling},
number = {3},
pages = {1136--1149},
pmid = {33682402},
title = {{Bayesian Particle Instance Segmentation for Electron Microscopy Image Quantification}},
volume = {61},
year = {2021}
}

@article{Smith2007,
author = {Smith, Ray},
doi = {10.1109/ICDAR.2007.4376991},
isbn = {0769528228},
issn = {15205363},
journal = {Proceedings of the International Conference on Document Analysis and Recognition, ICDAR},
pages = {629--633},
title = {{An overview of the tesseract OCR engine}},
volume = {2},
year = {2007}
}

@article{Liaw2018,
archivePrefix = {arXiv},
arxivId = {1807.05118},
author = {Liaw, Richard and Liang, Eric and Nishihara, Robert and Moritz, Philipp and Gonzalez, Joseph E. and Stoica, Ion},
eprint = {1807.05118},
number = {2012},
title = {{Tune: A Research Platform for Distributed Model Selection and Training}},
url = {http://arxiv.org/abs/1807.05118},
year = {2018}
}

@article{Dong2016,
archivePrefix = {arXiv},
arxivId = {1501.00092},
author = {Dong, Chao and Loy, Chen Change and He, Kaiming and Tang, Xiaoou},
doi = {10.1109/TPAMI.2015.2439281},
eprint = {1501.00092},
issn = {01628828},
journal = {IEEE Transactions on Pattern Analysis and Machine Intelligence},
number = {2},
pages = {295--307},
pmid = {26761735},
publisher = {IEEE},
title = {{Image Super-Resolution Using Deep Convolutional Networks}},
volume = {38},
year = {2016}
}

@article{Bochkovskiy2020,
archivePrefix = {arXiv},
arxivId = {2004.10934},
author = {Bochkovskiy, Alexey and Wang, Chien-Yao and Liao, Hong-Yuan Mark},
eprint = {2004.10934},
title = {{YOLOv4: Optimal Speed and Accuracy of Object Detection}},
url = {http://arxiv.org/abs/2004.10934},
year = {2020}
}

@article{He2016,
archivePrefix = {arXiv},
arxivId = {1512.03385},
author = {He, Kaiming and Zhang, Xiangyu and Ren, Shaoqing and Sun, Jian},
doi = {10.1109/CVPR.2016.90},
eprint = {1512.03385},
isbn = {9781467388504},
issn = {10636919},
journal = {Proceedings of the IEEE Computer Society Conference on Computer Vision and Pattern Recognition},
pages = {770--778},
publisher = {IEEE},
title = {{Deep residual learning for image recognition}},
volume = {2016-Decem},
year = {2016}
}

@article{Kim2020,
author = {Kim, Hyojin and Han, Jinkyu and Han, T. Yong Jin},
doi = {10.1039/d0nr04140h},
issn = {20403372},
journal = {Nanoscale},
number = {37},
pages = {19461--19469},
pmid = {32960204},
publisher = {Royal Society of Chemistry},
title = {{Machine vision-driven automatic recognition of particle size and morphology in SEM images}},
volume = {12},
year = {2020}
}

@article{Paszke2019,
archivePrefix = {arXiv},
arxivId = {1912.01703},
author = {Paszke, Adam and Gross, Sam and Massa, Francisco and Lerer, Adam and Bradbury, James and Chanan, Gregory and Killeen, Trevor and Lin, Zeming and Gimelshein, Natalia and Antiga, Luca and Desmaison, Alban and K{\"{o}}pf, Andreas and Yang, Edward and DeVito, Zach and Raison, Martin and Tejani, Alykhan and Chilamkurthy, Sasank and Steiner, Benoit and Fang, Lu and Bai, Junjie and Chintala, Soumith},
eprint = {1912.01703},
issn = {10495258},
journal = {Advances in Neural Information Processing Systems},
number = {NeurIPS},
title = {{PyTorch: An imperative style, high-performance deep learning library}},
volume = {32},
year = {2019}
}

@article{Schwenker,
author = {Schwenker, Eric and Jiang, Weixin and Spreadbury, Trevor and Ferrier, Nicola and Cossairt, Oliver and Chan, Maria K Y},
pages = {1--17},
title = {{EXSCLAIM ! – An automated pipeline for the construction of labeled materials imaging datasets from literature}}
}

@article{Redmon2017,
archivePrefix = {arXiv},
arxivId = {1612.08242},
author = {Redmon, Joseph and Farhadi, Ali},
doi = {10.1109/CVPR.2017.690},
eprint = {1612.08242},
isbn = {9781538604571},
journal = {Proceedings - 30th IEEE Conference on Computer Vision and Pattern Recognition, CVPR 2017},
pages = {6517--6525},
title = {{YOLO9000: Better, faster, stronger}},
volume = {2017-Janua},
year = {2017}
}

@article{Mukaddem2020,
author = {Mukaddem, Karim T. and Beard, Edward J. and Yildirim, Batuhan and Cole, Jacqueline M.},
doi = {10.1021/acs.jcim.9b00734},
issn = {15205142},
journal = {Journal of Chemical Information and Modeling},
number = {5},
pages = {2492--2509},
pmid = {31714792},
title = {{ImageDataExtractor: A Tool to Extract and Quantify Data from Microscopy Images}},
volume = {60},
year = {2020}
}

@article{Kononova2019,
    title = {Text-mined dataset of inorganic materials synthesis recipes},
    author = {Kononova, Olga and Huo, Haoyan and He, Tanjin and Rong, Ziqin and Botari, Tiago and Sun, Wenhao and Tshitoyan, Vahe and Ceder, Gerbrand},
    year = 2019,
    volume = 6,
    number = {203},
    journal = {Sci Data},
    doi = {10.1038/s41597-019-0224-1},
    url = {https://doi.org/10.1038/s41597-019-0224-1}
}

\section{Figures \& Tables}

\begin{table}[H]
\centering
\begin{tabularx}{\linewidth}{ L | L } 
 \textbf{Category} & \textbf{Keywords} \\
 \hline
 1 & TEM, SEM \\ 
 2 & Gold \\
 3 & sphere, cube, rod, triangle, prism, aunr \\
 \hline
\end{tabularx}
\caption{\textbf{Regular Expression Keywords corresponding to each category, utilized as a filter in the Paper Parsing stage.} To improve the precision of the collected figures, only those figures whose captions match at least one keyword from each of the three categories, are downloaded. Stemming is utilized to include grammatical variants of each keyword.}
\label{table:regex}
\end{table}

\begin{table}[H]
\centering
\begin{tabular}{ m{6.5em} | m{2cm} | m{2cm} | m{2cm} | m{3cm} } 
 \ & \textbf{Classifier-1} & \textbf{Classifier-2} & \textbf{Label, Scale and Bar Detector} & \textbf{Particle Segmentation}\\
 \hline
 \textbf{Task} & Image Classification & Image Classification & Object Detection & Instance Segmentation + Object Classification \\ 
 \hline
 \textbf{Classification Classes} & Microscopy, Non-Microscopy & Particulate, Non-Particulate & Label, Scale, Bar & Rod, Sphere, Cube, Triangle \\ 
 \hline
 \textbf{Dataset Size(no. of images)} & 1425 & 383 & 640 & 426\\ 
 \hline
 \textbf{Metric Score on Test Data } & 0.98 (F1 score) & 0.91 (F1 score) & 0.88 (mAP@50) & 0.77 (mAP@50)\\ 
 \hline
\end{tabular}
\caption{\textbf{Illustration of datasets and models used for training all machine learning models in the pipeline.} All datasets were divided into train, validation and test splits before model training. }
\label{table:classifier}
\end{table}

\begin{table}[H]
\centering
\begin{tabular}{ m{10em} | m{3.8cm} | m{2.5cm} | m{2.5cm} } 
\textbf{Data Description} & \textbf{Data Key Label} & \textbf{Data Type} & \textbf{Nested Keys and Types} \\
\hline
DOI of the parent paper & \texttt{DOI} & \emph{string} & \_ \\
\hline
A unique identifier of the image & \texttt{Hash} & \emph{string} & \_ \\
\hline
URL of the parent composite figure & \texttt{Composite\_Figure\_URL} & \emph{string} & \_ \\
\hline
Title of the parent composite figure & \texttt{Composite\_Figure\_Title} & \emph{string} & \_ \\
\hline
Morphology class of majority of the particles in the image & \texttt{Main\_class} & \emph{string} & \_ \\
\hline
Morphology classes that are present in lower frequency & \texttt{Minority\_classes} & \emph{list of strings} & \_ \\
\hline
Morphology-specific size measurements of all identified particles and units & \texttt{Size} & 
Object(\emph{dict}) &
\texttt{Unit}: \emph{string}
\texttt{Measurement}: Object(\emph{dict})$^a$ \\
\hline
Sub-figure label associated with the image & \texttt{Label} & \emph{string} & \_ \\
\hline
Information read from the scale & \texttt{Scale} & 
Object(\emph{dict}) & 
\texttt{digit}: \emph{string} 
\texttt{unit}: \emph{string} 
\texttt{bar\_length}: \emph{string} \\
\hline
\end{tabular}
\caption{\textbf{Illustration of the dataset schema.} \texttt{DOI}, \texttt{Hash} and \texttt{Composite\_Figure\_URL} are metadata fields that give general information about the source and nature of the image. \texttt{Main\_class}, \texttt{Minority\_class} and \texttt{Size} are fields that have been populated by the image analysis steps of the pipeline. The \texttt{Unit} and \texttt{Measurement} sub-fields of the \texttt{Size} field contain the measurement unit (nanometer or micron) and magnitudes of size measurements respectively. The \texttt{digit}, \texttt{bar\_length} and \texttt{unit} sub-fields of the \texttt{Scale} field contain the magnitude of the scale reading, horizontal length of the measurement bar and unit of measurement respectively. Schema of Object(\emph{dict})$^a$ is shown in Table ~\ref{table:measurement}. }
\label{table:full_schema}
\end{table}

\begin{table}[H]
\centering
\begin{tabularx}{1.0\linewidth}{ L | L | L } 
 \textbf{Data Key Label} & \textbf{Data Type} & \textbf{Nested Key and Type} \\
 \hline
 \texttt{rod} & Object(\emph{dict}) & 
 \texttt{length}: \emph{list of floats}
 \texttt{width}: \emph{list of floats} \\
 \hline
 \texttt{sphere} & Object(\emph{dict}) & 
 \texttt{diameter}: \emph{list of floats} \\
 \hline
 \texttt{triangle} & Object(\emph{dict}) & 
 \texttt{height}: \emph{list of floats} \\
 \hline
 \texttt{cube} & Object(\emph{dict}) & 
 \texttt{side}: \emph{list of floats} \\
\end{tabularx}
\caption{\textbf{Schema of the object linked to the Measurement sub-field.} It contains separate nested objects for each of the four morphologies. Each nested object contains fields with the measured dimension values.}
\label{table:measurement}
\end{table}

\end{document}